\documentclass[%
 preprint,
 singleocolumn,
 amsmath,amssymb,
 aps,
 pre,
showpacs]{revtex4-1}

\usepackage{graphicx}% Include figure files
\usepackage{dcolumn}% Align table columns on decimal point
\usepackage{bm}% bold math
\usepackage{subfigure}
\usepackage{float}
\usepackage{color}
\usepackage{array}
\usepackage{float}
\usepackage{multirow}

\begin{document}

%\preprint{APS/123-QED}

\title{Asymmetricity and sign reversal of secondary Bjerknes force from strong nonlinear coupling in cavitation bubble pairs}% Force line breaks with \\

\author{Vikash Pandey}
\email{Electronic address: vikashp@ifi.uio.no}
\affiliation{%
Centre for Ecological and Evolutionary Synthesis (CEES), Department of Biosciences, University of Oslo, P.O. Box 1066, NO-0316, Oslo, NORWAY\footnote[1]{Current affiliation.}
\\
 Research Centre for Arctic Petroleum Exploration (ARCE\textsc{x}), Department of Geosciences, UiT The Arctic University of Norway in Troms\o, P.O. Box 6050, N-9037, Troms\o, NORWAY
}%

\date{\today}% It is always \today, today,

\begin{abstract}
Most of the current applications of acoustic cavitation use bubble clusters that exhibit multibubble dynamics. This necessitates a complete understanding of the mutual nonlinear coupling between individual bubbles. In this study, strong nonlinear coupling is investigated in bubble pairs which is the simplest case of a bubble-cluster. This leads to the derivation of a more comprehensive set of coupled Keller-Miksis equations (KMEs) that contain nonlinear coupling terms of higher order. The governing KMEs take into account the convective contribution that stems from the Navier-Stokes equation. The system of KMEs is numerically solved for acoustically excited bubble pairs. It is shown that the higher order corrections are important in the estimation of secondary
Bjerknes force for closely spaced bubbles. Further, asymmetricity is witnessed in both magnitude and sign reversal of the secondary Bjerknes force in weak, regular, and strong acoustic fields. The obtained results are examined in the light of published scientific literature. It is expected that the findings reported in this paper may have implications
in industries where there is a requirement to have a control on cavitation and its effects.
\begin{description}

\item[PACS numbers]

\pacs{47.55.dd, 43.35.Ei, 47.55.Bx, 43.30.Nb, 43.25.Ts, 47.61.Jd, 43.25.Yw, 43.25.+y \vspace{20pt}  \\ \texttt{\large The peer-reviewed version of this manuscript is published in
\vspace{08pt}\\ 
\textit{Physical Review E, Vol.~99, No.~4,
Pages 042209, 2019.\vspace{08pt}\\
DOI: https://doi.org/10.1103/PhysRevE.99.042209}\vspace{08pt}\\
The published version of the manuscript is available online at\vspace{08pt}\\
\href{https://link.aps.org/doi/10.1103/PhysRevE.99.042209}{\textit{https://link.aps.org/doi/10.1103/PhysRevE.99.042209}}
\vspace{10pt}\\
\textcolor{blue}{This document is an e-print which may differ in, e.g.~pagination, referencing styles, figure sizes, and typographic details.} } }

% 47.55.dd Bubble dynamics
% 43.35.Ei Acoustic cavitation in liquids
% 47.55.Bx Cavitation
% 43.25.Ts Nonlinear acoustical and dynamical systems\\
% 43.30.Nb Noise in water; generation mechanisms and characteristics of the field
% 47.61.Jd Multiphase flows
% 43.25.Yw Nonlinear acoustics of bubbly liquids
% 43.35.Hl Sonoluminescence
% 05.45.−a Nonlinear dynamics and chaos
% 43.25.+y Nonlinear acoustics
% 47.35.Rs Sound waves

\end{description}
\end{abstract}


\maketitle

%\tableofcontents

\section{\label{sec:level1}Introduction}

Acoustic cavitation is defined as the formation and pulsation of gas
cavities in a liquid under the action of an acoustic field \cite{Wu2017,Zilonova2018}.
The cavities often grow as microbubbles and exhibit a cache of exotic phenomena in their relatively short life-time. This may include; rapid
oscillations, high speed liquid micro-jets, emanation of capillary
waves and shock waves, and finally a violent collapse with sonoluminescence
\cite{Lauterborn2010,Koukouvinis2016,Lai2018}. The motivation to
study cavitation bubbles has changed in the last one century from
scientific curiosity \cite{Rayleigh1917} to application driven \cite{Weinberg2009,Kooiman2011,Rabaud2011,Maksimov2014,Verhaagen2016,Zilonova2018}. In most applications, bubble-clusters are manipulated by irradiating them with an acoustic wave. Bubbles first experience a primary Bjerknes force due to the direct impact from the incident wave. Consequently, bubbles may translate toward pressure antinodes if their equilibrium radii is smaller than the resonant size corresponding to the wave frequency. Otherwise, bubbles migrate toward pressure nodes. 

But, this study focuses on the secondary Bjerknes force that oscillating bubbles exert on each other through the re-radiation of the acoustic field \cite{Crum1975}. This is because even though the primary Bjerknes force is usually stronger than the secondary Bjerknes force for bubbles separated by large distances, the former can be successfully circumvented through clever designing of the experimental apparatus \cite{Crum1975,Yoshida2011}. Besides, the origin and implications of the primary Bjerknes force are quite well understood. Further, recent applications of cavitation involve closely spaced bubbles in strong driving fields in which secondary Bjerknes force dominates over the primary Bjerknes force \cite{Jiao2015a,Han2015}.

The secondary Bjerknes force is attractive when bubbles oscillate in phase, and repulsive for out of phase oscillations. According to the linear Bjerknes theory, repulsion occurs when the frequency of the driving acoustic field lies between the Minnaert (linear-resonance) frequencies of the two bubbles \cite{Crum1975}. The Bjerknes force is attractive otherwise. The sign reversal of the secondary Bjerknes force from attraction to repulsion is quite rare. It was experimentally confirmed more than a decade after its theoretical prediction \cite{Barbat1999}. Zabolotskaya predicted the sign reversal using the linear Bjerknes theory, and attributed it to the change in the oscillation frequency of the bubbles due to their mutual interaction \cite{Zabolotskaya1984}. Similar results were also obtained by Oguz and Prosperetti but using nonlinear theory \cite{Oguz1990}. However, their theory could not predict the formation of \textit{bubble-grape} like stable structures if bubbles larger than the resonant sizes were closely-spaced in a weak acoustic field. This was contrary to the experimental observations \cite{Doinikov1995b}. 

In contrast, Pelekasis and Tsamopoulos included shape deformations due to subharmonic resonances, and predicted repulsive forces for closely-spaced asymmetric bubble-pairs \cite{Pelekasis1993b}. Even though their result was in agreement with the linear theory, they attributed the sign reversal to the nonlinear coupling between the bubbles. Ida showed that sign reversal could occur in asymmetric bubble-pairs due to the second-highest transition frequency of the smaller bubble \cite{Ida2003}. Doinikov took multiple scattering into account and described the sign reversal due to change in effective resonance frequencies of the bubbles \cite{Doinikov1995a}. The change is because of the stiffening that arises when bubbles oscillate in phase with each other. The second-order harmonics generated in strong acoustic fields may also yield a sign reversal \cite{Doinikov1999}. Doinikov further used Lagrangian formalism to obtain sign reversal in bubble-pairs separated by large distances in a strong driving field \cite{Doinikov2004a}. In comparison to Doinikov's approach of including third-order terms, Harkin showed the importance of fourth order terms in the generation of radial harmonics and nonlinear phase shifting, that leads to the sign reversal \cite{Harkin2001}. Clearly, this indicates the importance of higher order terms in the study of bubble dynamics. 

An alternative mechanism for sign reversal based on the linear theory was recently presented in Ref.~\cite{Lanoy2015} which seems similar to that from Feuillade \cite{Feuillade1995}. There, the coupled bubble system is shown to have two modes of oscillations: symmetric and asymmetric. Contrary to the symmetric mode that dominates at a resonance close to the resonance frequency of the larger bubble, the antisymmetric mode comes into play at a resonance close to the resonance frequency of the smaller bubble. The symmetric and antisymmetric modes give rise to attractive secondary Bjerknes force and repulsive secondary Bjerknes force respectively. 

Thus, it can be inferred that the nonlinearity inherent in bubble dynamics make them a subject difficult
for investigation \cite{Sugita2017b,Zhang2018}. This can be ascertained from the fact that the study of bubble dynamics has benefited chaos
theory, and vice-versa \cite{Lauterborn1981,Lauterborn1988,Zhang2018}. Bubble-systems are known to display chaotic oscillations through period doubling \cite{Lauterborn2010, Louisnard2011}. Similarities between bifurcation diagrams from different bubble-systems have been observed if the ratio of the equilibrium radius to the wavelength of the sound wave is same in them \cite{Behnia2009}. Interestingly, the presence of electrical charges on bubbles may advance the bifurcations \cite{Hongray2015}. But, a bubble can only carry a finite charge since there is a lower limit on the radius that it could attain during its collapse phase. Since the charges on a bubble reduces its surface tension, a charged bubble may grow to a relatively larger radius, and also contract to a smaller radius. Charged bubbles also have a greater collapse velocity and a lower Blake threshold \cite{Hongray2014}. Poincar\'{e} maps have further confirmed that bubbles that are slightly smaller than the resonant size oscillates chaotically in both radial and translational directions \cite{Watanabe1993}. Though, the chaos observed in the translational direction is attributed to the nonlinear radial oscillations. An attractor from chaotic bubble dynamics may exhibit fractality too \cite{Louisnard2011}.

The goal of this paper is to investigate strong nonlinear coupling
in a two-bubble system which is the simplest case of a bubble
cluster. The motivation behind this study is three-fold. First, cavitation
bubbles do not occur in isolation, but rather they come in ensembles
or clusters. Such clusters are characterized by multi-bubble dynamics which makes
their modeling difficult, both analytically and computationally \cite{Ma2018}.
The possibility of a bubble cluster to originate from a single bubble \cite{Rabaud2011},
and multibubbles coalescing to form a single bubble, indicates the importance
of nonlinear coupling between bubbles \cite{Pishchalnikov2011}. Besides, the nonlinear interactions in a bubble cloud could lower the subharmonic resonance frequency as well as the corresponding minimum pressure threshold that leads to the emission of such subharmonics \cite{Guedra2017}. On
the one hand, nonlinear coupling between bubbles has been ignored
by approximating a bubble cluster as a single bubble \cite{Chahine1984}. On the other hand, this may not be valid at high frequencies, and in narrow bubble size distributions as then there is a significant increase in the individual scattering cross-section \cite{Feuillade1995}. Besides, the coupling effects between bubbles separated by small
distances in strong acoustic fields may be significant enough to
alter the behavior of the whole cluster \cite{Kooiman2014}. Unfortunately, the nonlinearity in bubble-dynamics has mostly been approximated by the use of linear theories \cite{Pelekasis1993a,Sugita2017b}. The inadequacy of such an approach has already been accepted by the scientific
community \cite{Ilinskii1992,Pishchalnikov2011,Yoshida2011,Maksimov2016}. 
Alternatively, coupling effects are ignored by assuming bubbles to
be separated by a distance much larger than their individual sizes
\cite{Chahine1992}. In such cases, linear superposition of the responses from the individual bubbles is considered identical to the collective response of the bubble cluster. But this may not be valid for many applications \cite{Ilinskii1992}.

The coupling between the bubbles has also been known to affect the bubble-size distribution within a cluster \cite{Yasui2009} that necessitated the use of statistical techniques for their investigation \cite{Glimm1990}. Since, bubble-clusters facilitate a
greater number of reaction sites, an understanding of bubble-bubble
interactions may give an insight into the factors that lead to the
difference in temperature that is observed in sonoluminescence from
a single bubble and multibubbles \cite{Matula1995}. It has also been experimentally shown that the secondary Bjerknes force favors the formation of bubble-clusters instead of dendritic filament branches which alters the sonoluminescence intensity \cite{Hatanaka2002}. Thus, the study of multibubble dynamics may benefit
sonochemical engineering applications where quenching of sonoluminescence
is frequently encountered \cite{Didenko2000,Cairos2017,Zhang2018}.
Moreover, the coupling between the individual bubbles in a collapsing bubble cloud may play an important role in the generation of shock waves and broadband cavitation noise \cite{Ida2009a,Yasui2010,Song2016,Liang2017}.
Such bubble-clouds are regularly generated from ship propellers, and from the
firing of air guns in marine seismic explorations \cite{Graaf2014a,Aktas2016}.
A complete understanding of the nonlinear coupling is essential in the designing
of efficient propellers and gun-arrays as they have direct implications
for underwater navigation and communication.

Second, some recent experimental studies have shown that bubbles of
different sizes when separated by a small distance exhibit strong nonlinear coupling \cite{Ochiai2015,Liang2017,Ma2018}.
Such a case was examined in Ref.~\cite{Maksimov2016} using bi-spherical
coordinates, but that was restricted to small amplitude oscillations. A relatively successful theory of nonlinear coupling
between bubbles was presented in Ref.~\cite{Mettin1997} almost twenty
years ago which also motivates the study presented in this paper. However,
Keller-Miksis equations (KMEs) derived to describe the bubble-bubble
interaction in Ref.~\cite{Mettin1997} were based on assumptions that
could be valid only for weak coupling. That would be plausible either
under the action of a weak acoustic field, or,
when bubbles are far apart from each other. An evidence of weak coupling
can be seen from Fig.~5(b) in Ref.~\cite{Mettin1997} where the radius
evolution of the larger bubble remains unaffected from the smaller
bubble at all times. Such an assumption of weak coupling was reasonable
for applications twenty years ago. But modern applications involve bubbles in immediate vicinity of each other. Some examples can be seen in Fig.~12 in Ref.~\cite{Weinberg2009} for shock wave lithotripsy, Fig.~2 in Ref.~\cite{Kooiman2011} for sonoporation, Fig.~2 in Ref.~\cite{Rabaud2011} for micorfluidics, and Fig.~4 in Ref.~\cite{Verhaagen2016} for cleaning. Moreover, most equations proposed to study single bubble dynamics and multibubble
dynamics are modifications of the Rayleigh-Plesset equation that takes
into account the convective contribution from the Navier-Stokes equation (NSE) in its derivation \cite{Rayleigh1917,Keller1956,Ilinskii1992}.
But, the theories that are currently used to study the radiation-coupling in bubbles seem to have ignored this because of the analytic
and computational complexities that come along with the higher order
terms. Thus, contrary to the relevance of the higher order terms in cavitation-like
highly nonlinear dynamic process \cite{Doinikov2001,Harkin2001},
they were ignored in Ref.~\cite{Mettin1997} too. Thus, in light of this
observation, the theory from Ref.~\cite{Mettin1997} will be regarded
as the theory of weak nonlinear coupling in this paper. 

Third, there is no clear consensus on what leads to the asymmetricity in the secondary Bjerknes
force. A time-delay in radiation due to the large separation between the
bubbles has been predicted as one of the possible causes for  it in Ref.~\cite{Mettin1997}. Another possibility
as mentioned in Ref.~\cite{Pelekasis2004} is asymmetric viscous dissipation
of the forces if bubbles are of different size. But, since the fluid
was assumed to be inviscid in that study, the authors had themselves admitted
it to be counter-intuitive.

The rest of the article is organized as follows. Sec.~\ref{sec:Nonlinear-coupling-between} is divided into two subsections.
In its first subsection, the established theory of weak nonlinear coupling \cite{Mettin1997}, is briefly
summarized. In the second subsection, the theory of strong nonlinear coupling is developed
and subsequently a system of strongly coupled KMEs is derived. The numerical solutions
of the KMEs are presented in Sec.~\ref{sec:Numerical-results} where they are also compared with the respective solutions from the weak nonlinear coupling. Finally, in Sec.~\ref{sec:Conclusion}, the implications of this work is discussed.

\section{Nonlinear coupling between a pair of bubbles\label{sec:Nonlinear-coupling-between}}

It is assumed that two cavitation-bubbles with volumes $V_{1}$ and
$V_{2}$, are formed in an incompressible fluid of density,
$\rho$, and dynamic viscosity, $\mu$. The two bubbles referred as $B1$ and $B2$, have the radii $R_{1}\left(t\right)$ and $R_{2}\left(t\right)$
respectively, and are separated by a distance, $d$, that is measured
from the bubble centers. The radial velocities of the bubble boundaries
at any given time, $t$, are then expressed as, $\dot{R}_{1}\left(t\right)$
and $\dot{R}_{2}\left(t\right)$, where the number of over-dots represent
the order of differentiation with respect to time. So, a double-dot,
and a triple-dot over $R$ would imply an acceleration and \textit{jerk}
of the bubble boundaries respectively. Similar to Refs.~\cite{Mettin1997,Maksimov2016}, the bubbles are
assumed spherical in shape which is valid for bubbles smaller than
the wavelength of the acoustic wave that excites it. So, scattering
effects are neglected too. Also, thermal and viscous dissipation are ignored.

The fluid flow during cavitation can safely be assumed to be radially symmetric. Therefore, the
continuity equation in the spherical coordinate system contains contribution only from the radial-components
of the flow, none from azimuthal and polar parts. Consequently, 
\begin{equation}
u_{1}=\left(\frac{R_{1}}{r}\right)^{2}\dot{R}_{1}\label{vel_eq},
\end{equation}
where $r$ is the distance from the center of the $B1$, and
$u_{1}$ is the radial velocity of the fluid in the immediate vicinity
of the boundary of $B1$. Further, the nonlinear coupling theory developed in this
paper builds on the modified KME \cite{Mettin1997}, which is expressed
for $B2$ as:
\begin{equation}
\left(1-\frac{\dot{R}_{2}}{c}\right)R_{2}\ddot{R}_{2}+\frac{3}{2}\dot{R}_{2}^{2}\left(1-\frac{\dot{R}_{2}}{3c}\right)=\left(1+\frac{\dot{R}_{2}}{c}\right)\frac{P}{\rho}+\frac{R_{2}}{\rho c}\frac{dP}{dt},\label{main_kme}
\end{equation}
where $P=P_{\text{bw}}-P_{\text{stat}}+P_{\text{v}}-P_{\text{ext}}$. The liquid pressure at the bubble wall, $P_{\text{bw}}$, is expressed as
\begin{equation}
\text{ }P_{\text{bw}}=\left(P_{\text{stat}}-P_{\text{v}}+\frac{2\sigma}{R_{20}}\right)\left(\frac{R_{20}}{R_{2}}\right)^{3\gamma}-\frac{2\sigma}{R_{2}}-4\mu\frac{\dot{R}_{2}}{R_{2}},\label{bubble_wall_pressure}
\end{equation}
where $R_{20}$ and $\sigma$ are the equilibrium radius and surface tension of $B2$ respectively. Further, $c$ is the sound speed in water, $\gamma$ is the polytropic index of the gas inside $B2$, $P_{\text{stat}}$ is the hydrostatic pressure far away from $B2$, $P_{\text{v}}$ is the vapor pressure, and $P_{\text{ext}}=P_{s}\sin\left[2\pi f_{s}\left(t+(R_{2}/c)\right)\right]$ is the time-delayed external driving acoustic field of amplitude, $P_{s}$, and frequency, $f_{s}$. 

\subsection{Weak nonlinear coupling\label{subsec:Weak-nonlinear-coupling}}

The framework presented to study weak nonlinear coupling in bubbles in Ref.~\cite{Mettin1997}, is summarized here. If $p_{1}$ is the pressure set-up by an oscillating bubble, $B1$,
around its surrounding fluid, then from Eq.~$\left(3\right)$ in Ref.~\cite{Mettin1997}, we have 
\begin{equation}
p_{1}=\frac{\rho}{r}\left(2R_{1}\dot{R}_{1}^{2}+R_{1}^{2}\ddot{R}_{1}\right).\label{weak_pressure}
\end{equation}
The radiation force exerted by B1 on B2 is (see Eq.~$\left(5\right)$
in Ref.~\cite{Mettin1997})
\begin{equation}
F_{12}^{w}=\frac{\rho V_{2}}{4\pi d^{2}}\frac{d^{2}V_{1}}{dt^{2}}\hat{e}_{r},\label{weak_bjerknes12}
\end{equation}
where, $\hat{e}_{r}$ denotes the radial unit vector, and the superscript
``$w$'' symbolizes the weak nonlinear coupling aspect
of the force. The secondary Bjerknes force, $F_{B}^{w}$, is calculated
by time-averaging the radiation force as (see Eq.~$\left(6\right)$
in Ref.~\cite{Mettin1997})
\begin{equation}
F_{B}^{w}=\left\langle F_{12}^{w}\right\rangle =-\frac{\rho}{4\pi d^{2}}\left\langle \dot{V_{1}}\dot{V_{2}}\right\rangle \hat{e}_{r},\label{weakBjerknes_force}
\end{equation}
where  $\left\langle \cdot \right\rangle$ denotes time-averaging over a period of the incident wave. The attractive and repulsive nature of the time-averaged secondary Bjerknes force are respectively represented by their negative and positive signs. It can be seen from
Eq.~$\left(\ref{weakBjerknes_force}\right)$ that interchanging the indices $1\leftrightarrow2$
does not change the expression on the right hand side. Thus, the secondary Bjerknes force is symmetric in the same coordinate system as
\begin{equation}
\left\langle F_{12}^{w}\right\rangle =-\left\langle F_{21}^{w}\right\rangle .\label{symmetry_bjerknes}
\end{equation}
Replacing $P_{\text{ext}}$ in Eq.~$\left(\ref{main_kme}\right)$ by $P_{\text{ext}}+p_{1}$, and then substituting Eq.~$\left(\ref{weak_pressure}\right)$ in it, the governing KME for $B2$ is obtained as (see Eq.~$\left(07\right)$
in Ref.~\cite{Mettin1997})
\begin{equation}
\left(1-\frac{\dot{R}_{2}}{c}\right)R_{2}\ddot{R}_{2}+\frac{3}{2}\dot{R}_{2}^{2}\left(1-\frac{\dot{R}_{2}}{3c}\right)=\left(1+\frac{\dot{R}_{2}}{c}\right)\frac{P_{\text{bw}}-P_{\text{stat}}+P_{\text{v}}-P_{\text{ext}}}{\rho}+\frac{R_{2}}{\rho c}\frac{d}{dt}\left[P_{\text{bw}}-P_{\text{ext}}\right]-C_{2}^{w},\label{mettin_kme}
\end{equation}
where the \textit{coupling dose, $C_{2}^{w}$,} received by $B2$
from $B1$ is 
\begin{equation}
C_{2}^{w}=\frac{1}{d}\left(2R_{1}\dot{R}_{1}^{2}+R_{1}^{2}\ddot{R}_{1}\right).\label{mettin_dose}
\end{equation}
The motivation to call the last term in Eq.~$\left(\ref{mettin_dose}\right)$
as coupling dose is due to its units, $J/kg$.

\subsection{Strong nonlinear coupling\label{subsec:Strong-nonlinear-coupling}}
The Navier-Stokes equation for a purely radial flow due to the pulsations of $B1$ is
\begin{equation}
\frac{Du_{1}}{Dt}\equiv\frac{\partial u_{1}}{\partial t}+u_{1}\frac{\partial u_{1}}{\partial r}=-\frac{1}{\rho}\frac{\partial p_{1}}{\partial r}+\frac{\mu}{\rho}\left[\frac{1}{r^{2}}\frac{\partial}{\partial r}\left(r^{2}\frac{\partial u_{1}}{\partial r}\right)-\frac{2u_{1}}{r^{2}}\right]+f_{\text{ext}},\label{mat_deriv}
\end{equation}
where $D/Dt$ is the material derivative operator expressed as the
sum of an instantaneous local (time) derivative and a convective (spatial)
derivative, and $f_{\text{ext}}$ represents external
body forces such as gravity, Lorentz force, etc, that could be affecting
the flow. The spatial-derivative term describes the acceleration that one observes while moving radially along the velocity flow field, $u_{1}$. And, the time-derivative term describes the intrinsic variation of the field. It should be noted that contrary to the local acceleration which is zero for a steady flow, i.e., when $u_1$ is constant in time, the convective acceleration may be non-zero. Even though the convective contribution diminishes rapidly for most fluid flows in their far field, it may play a major role in the near field. Since, $u_{1}\cdot\partial u_{1}/\partial r=\partial u_{1}^{2}/2\partial r$, the convective term is the main source of nonlinearity that may also lead to chaos and turbulence. Therefore, it is only natural to include the convective contribution in order to explore the complete nonlinearity. 

Further, since shear viscosity does not generate any resistance in a purely radially symmetric flow around a bubble, the viscous term vanishes off completely from the Eq.~$\left(\ref{mat_deriv}\right)$. But, as shown
in Eq.~$\left(\ref{bubble_wall_pressure}\right)$, there is some viscous contribution that comes from the shear stress at the bubble
boundary.
The local and convective derivative terms of Eq.~$\left(\ref{mat_deriv}\right)$ are obtained by applying appropriate derivative operators on Eq.~$\left(\ref{vel_eq}\right)$. Substituting the derivative terms back in Eq.~$\left(\ref{mat_deriv}\right)$, and  assuming, $f_{\text{ext}}=0$, we get
\begin{equation}
\frac{1}{r^{2}}\left(2R_{1}\dot{R}_{1}^{2}+R_{1}^{2}\ddot{R}_{1}\right)-\frac{2R_{1}^{4}\dot{R}_{1}^{2}}{r^{5}}=-\frac{1}{\rho}\frac{\partial p_{1}}{\partial r}.\label{big_deriv}
\end{equation}
Integrating Eq.~$\left(\ref{big_deriv}\right)$ on both sides with respect to $r$, gives
\begin{equation}
p_{1}=\rho\left[\frac{1}{r}\left(2R_{1}\dot{R}_{1}^{2}+R_{1}^{2}\ddot{R}_{1}\right)-\frac{R_{1}^{4}\dot{R}_{1}^{2}}{2r^{4}}\right].\label{strong_pressure}
\end{equation}
Interestingly, the importance of the two terms in Eq.~$\left(\ref{strong_pressure}\right)$
has been acknowledged in Ref.~\cite{Keller1956}, where the first term
that is inversely proportional to $r$ was found important near bubble minima
such that it accounts for the primary shock wave pressure and subsequent
pressure peaks. The second term that is inversely proportional to the fourth power in $r$
was termed as the ``\textit{afterflow}'' pressure which is important
between pressure peaks at small distances from the bubble. It should be clarified that the afterflow term stems from the convective acceleration of the fluid. Unfortunately, this term was ignored in Eq.~$\left(\ref{weak_pressure}\right)$ to develop the theory of weak nonlinear coupling in Ref.~\cite{Mettin1997}. Though such an approach is quite acceptable for small amplitude oscillations \cite{Prosperetti1980}, convective contribution may become significant for large amplitude nonlinear oscillations where they may also give rise to microstreaming like second order effects \cite{Prosperetti1984}. The pressure gradient, $\partial p_{1}/\partial r$,
set-up radially outward by $B1$ leads to a radiation force, $F_{12}^{s}$, which is expressed as
\begin{equation}
F_{12}^{s}=-V_{2}\frac{\partial p_{1}}{\partial r}\Biggl|_{r=d}\hat{e}_{r},\label{strong_Bjerknes12}
\end{equation}
where the superscript ``$s$'' symbolizes the strong nonlinear coupling
aspect of the force. Extracting $\partial p_{1}/\partial r$ from
Eq.~$\left(\ref{big_deriv}\right)$ and substituting it in Eq.~$\left(\ref{strong_Bjerknes12}\right)$,
we get
\begin{equation}
F_{12}^{s}=\rho V_{2}\left[\frac{1}{d^{2}}\left(2R_{1}\dot{R}_{1}^{2}+R_{1}^{2}\ddot{R}_{1}\right)-\frac{2R_{1}^{4}\dot{R}_{1}^{2}}{d^{5}}\right]\hat{e}_{r},\label{strong_Bjerknes_complete}
\end{equation}
which can also be expressed in terms of bubble volumes as
\begin{equation}
F_{12}^{s}=\frac{\rho V_{2}}{4\pi d^{2}}\left[\frac{d^{2}V_{1}}{dt^{2}}-\frac{1}{2\pi d^{3}}\left(\frac{dV_{1}}{dt}\right)^{2}\right]\hat{e}_{r}.\label{final_strong_Bjerknes}
\end{equation}
The secondary Bjerknes force, $F_{B}^{s}$, is estimated by time-averaging $F_{12}^{s}$ from Eq.~$\left(\ref{final_strong_Bjerknes}\right)$ over the time-period, $\tau$, of the driving acoustic field as
\begin{equation}
\text{ }F_{B}^{s}=\left\langle F_{12}^{s}\right\rangle =\frac{\rho}{4\pi d^{2}}\left[I_{1}-I_{2}\right]\hat{e}_{r},\label{integral_eq}
\end{equation}
where 
\begin{equation}
I_{1}=\frac{1}{\tau}\int\limits _{t}^{t+\tau}V_{2}\left(t'\right)\frac{d^{2}V_{1}\left(t'\right)}{dt'^{2}}dt',\text{ }I_{2}=\frac{1}{2\pi d^{3}}\frac{1}{\tau}\int\limits _{t}^{t+\tau}V_{2}\left(t'\right)\left(\frac{dV_{1}\left(t'\right)}{dt'}\right)^{2}dt,\label{both_integrals}
\end{equation}
and $t'$ is a dummy variable for integration. Evaluating the integrals
in Eq.~$\left(\ref{both_integrals}\right)$, and substituting them back in Eq.~$\left(\ref{integral_eq}\right)$,
we get
\begin{equation}
F_{B}^{s}=\left\langle F_{12}^{s}\right\rangle =-\frac{\rho}{4\pi d^{2}}\left[\left\langle \dot{V_{1}}\dot{V_{2}}\right\rangle +\frac{1}{2\pi d^{3}}\left\langle \dot{V}_{1}^{2}V_{2}\right\rangle \right]\hat{e}_{r}.\label{avg_strong_Bjerknes}
\end{equation}
There are three main differences that can be observed on comparing Eqs.~$\left(\ref{final_strong_Bjerknes}\right)$ and $\left(\ref{avg_strong_Bjerknes}\right)$ that are obtained for the secondary Bjerknes force from strong nonlinear coupling, with the respective Eqs.~$\left(\ref{weak_bjerknes12}\right)$ and $\left(\ref{weakBjerknes_force}\right)$ obtained from the weak nonlinear coupling. First, the presence of the last term in Eq.~$\left(\ref{avg_strong_Bjerknes}\right)$ revokes the symmetry restrictions of the Bjerknes force as
\begin{equation}
\left\langle F_{12}^{s}\right\rangle \neq-\left\langle F_{21}^{s}\right\rangle.\label{asymmetric_Bjerknes}
\end{equation}
It will be further shown in the next section that the asymmetricity is observed in the magnitude as well as in the sign reversal of the Bjerknes force. An interesting finding is that $\left\langle F_{12}^{s}\right\rangle$ and $\left\langle F_{21}^{s}\right\rangle$ may not necessarily be equal even in a symmetric bubble-pair, i.e., when $V_{1}=V_{2}$. This is because identical bubbles may not necessarily have the same radial velocity if the coupling between them is nonlinear. But, the symmetricity of the Bjerknes force could be restored if the bubbles are separated by a sufficiently large distance, $d$, such that the convective contribution from the last term in Eq.~$\left(\ref{avg_strong_Bjerknes}\right)$ becomes negligible.

Second, contrary to most studies in which secondary Bjerknes force is assumed to have an inverse square law form along the lines of gravitational,
electrostatic, and magnetic force \cite{Crum1975}, it is observed that such an assumption may not be valid if there is a major contribution from the convective term in Eq.~$\left(\ref{avg_strong_Bjerknes}\right)$. This may happen in the case of closely spaced bubbles as then the bubbles are strongly coupled and influence each-other's oscillations significantly \cite{Ida2003,Lanoy2015}. Interestingly, this limitation of the linear Bjerknes theory was also \textit{subtly} mentioned by Crum in the paragraph following Eq.~(6) in Ref.~\cite{Crum1975}.

Third, the generalized observation made about the local and spatial derivative terms in the paragraph following Eq.~$\left(\ref{mat_deriv}\right)$ can also be verified from the expressions of the Bjerknes force. In the case when $B1$ has a constant velocity, i.e., $\dot u_{1}\left(t\right)=0$, then following the classical notion from Newton's second law, no force would be exerted from $B1$ to any other bubble in its neighborhood. This corresponds to the case of $\ddot V_{1}=0$, which yields $F_{12}^{w}=0$ from Eq.~$\left(\ref{weak_bjerknes12}\right)$. Further, $F_{21}^{w}=0$, due to the symmetry restriction from Eq.~$\left(\ref{symmetry_bjerknes}\right)$. But, contrary to this expectation from the classical mechanics, even in the case when $\ddot V_{1}=0$, the presence of the last term from the convective contribution in Eq.~$\left(\ref{final_strong_Bjerknes}\right)$ guarantees an attractive force, $F_{12}^{s}<0$, on $B2$ due to $B1$. Thus, as mentioned in Ref.~\cite{Ilinskii1992}, it may be difficult to \textit{directly} relate the convective contribution to the radiation field in a classical sense. Further, since $\dot V_{1}$ is squared in the last term in Eq.~$\left(\ref{final_strong_Bjerknes}\right)$, its contribution to the Bjerknes force always remains attractive at all instants of the driving sound field. Consequently, the convective (\textit{afterflow}) contribution adds to the attraction between the bubbles, but opposes their mutual repulsion.

The response of a bubble to the acoustic field emanated from another
bubble is similar in principle in the way it reacts to the external
driving pressure, $P_{\text{ext}}$. So, for $B2$, we have, $P=P_{\text{bw}}-P_{\text{stat}}+P_{\text{v}}-\left(P_{\text{ext}}+p_{1}\right)$,
where $p_{1}$ given by Eq.~$\left(\ref{strong_pressure}\right)$ is evaluated
at $r=d$. Substituting this expression for $P$ in Eq.~$\left(\ref{main_kme}\right)$,
we arrive at the final governing KME for $B2$ that encompasses the strong
nonlinear coupling effects from $B1$ as
\begin{equation}
\left(1-\frac{\dot{R}_{2}}{c}\right)R_{2}\ddot{R}_{2}+\frac{3}{2}\dot{R}_{2}^{2}\left(1-\frac{\dot{R}_{2}}{3c}\right)=\left(1+\frac{\dot{R}_{2}}{c}\right)\frac{P_{\text{bw}}-P_{\text{stat}}+P_{\text{v}}-P_{\text{ext}}}{\rho}+\frac{R_{2}}{\rho c}\frac{d}{dt}\left[P_{\text{bw}}-P_{\text{ext}}\right]-C_{2}^{s},\label{vikash_kme}
\end{equation}
where the strong coupling dose, $C_{2}^{s}$, can be grouped as a
sum of three terms; $C_{2}^{s}=C_{21}^{s}+C_{22}^{s}+C_{23}^{s}$, such that,
\begin{equation}
\text{\ensuremath{C_{21}^{s}=\frac{1}{d}\left(2R_{1}\dot{R}_{1}^{2}+R_{1}^{2}\ddot{R}_{1}\right)}},\label{vikash_dose1}
\end{equation}
\begin{equation}
C_{22}^{s}=\frac{1}{cd}\left(2\dot{R}_{2}R_{1}\dot{R}_{1}^{2}+\dot{R}_{2}R_{1}^{2}\ddot{R}_{1}+2R_{2}\dot{R}_{1}^{3}+6R_{2}R_{1}\dot{R}_{1}\ddot{R}_{1}+R_{2}R_{1}^{2}\dddot{R}_{1}\right),\label{vikash_dose2}
\end{equation}
\begin{equation}
\text{and }C_{23}^{s}=-\frac{1}{2cd^{4}}\left(cR_{1}^{4}\dot{R}_{1}^{2}+\dot{R}_{2}R_{1}^{4}\dot{R}_{1}^{2}+4R_{2}R_{1}^{3}\dot{R}_{1}^{3}+2R_{2}R_{1}^{4}\dot{R}_{1}\ddot{R}_{1}\right).\label{vikash_dose3}
\end{equation}
It can be seen that the term $C_{21}^{s}$ has the same expression as the coupling dose term, $C_{2}^{w}$, in Eq.~$\left(\ref{mettin_dose}\right)$ from weak nonlinear coupling. The respective KME for $B1$ can be obtained by interchanging the
indices $1\leftrightarrow2$ in the set of Eqs.~$\left(\ref{vikash_kme}\right)$--$\left(\ref{vikash_dose3}\right)$. The presence of the jerk term, $\dddot{R}_{1}$,
in Eq.~$\left(\ref{vikash_dose2}\right)$ is not new in the study of bubble dynamics.
On the one hand, it has been shown that a jerk equation is equivalent
to a system of three first order, ordinary, non-linear differential
equations which is a minimal setting for chaotic solutions \cite{Chlouverakis2006}. Thus, the observation of chaotic oscillations is not surprising in multi-bubble systems \cite{Lauterborn1988,Zhang2018}. On the other hand,
the jerk-term also invokes Ostrogradsky's instability \cite{Motohashi2015}.

\section{Numerical results and discussion\label{sec:Numerical-results}}

It is assumed that the gas bubbles undergo adiabatic oscillations in water. So, the numerical values used for the physical variables are: $P_{\text{stat}}=1\text{ }bar=101.325\text{ kPa}$,
$P_{\text{v}}=2.3388\text{ kPa}$, $\sigma=0.0728\text{ }N/m$,
$\rho=998.207\text{ }kg/m^{3}$, $\mu=0.001002\text{ }kg/(ms)$, $c=1481\text{ }m/s$, and $\gamma=1.4$. The jerk-term in Eq.~$\left(\ref{vikash_dose2}\right)$ is neglected to circumvent the Ostrogradsky instability in numerical solutions. An approximation of the jerk term could be possible \cite{Ilinskii1992,Maksimov2014,Maksimov2016}, but it is avoided because of the skewedness from the approximation error that leads to the non-convergence of the numerical algorithms which are
used to solve the KMEs \cite{Cox2014}.

The set of coupled KMEs is solved for both weak nonlinear coupling and strong nonlinear coupling for three bubble-pairs that are under excitation from acoustic fields of different strengths and frequencies. The bubble-pairs are: $\left(6,5\right)$,
$\left(10,5\right)$, and $\left(112,22\right)$, where the two numeric
values inside the parentheses are the equilibrium radii of $B1$, and
$B2$, in $\mu m$, respectively. Such a choice for bubble-pairs is driven by two reasons. First, the pairs $\left(6,5\right)$ and $\left(10,5\right)$ have also been studied in Ref.~\cite{Mettin1997}. Therefore, a direct comparison of the results from the weak coupling and the strong coupling would be possible. And, since the behavior of the third pair $\left(112,22\right)$ has been observed against the predictions from the linear Bjerknes theory in experiments \cite{Yoshida2011}, this may further provide a validation of the results from strong nonlinear coupling. Besides Ref.~\cite{Yoshida2011}, there is experimental observation of repulsive secondary Bjerknes force in Ref.~\cite{Lanoy2015} too, but the data from the latter cannot be used for validation. This is because the pressure reported in Ref.~\cite{Lanoy2015} is in volts which cannot be directly converted into pascals due to the complexity of the experimental procedure. The second reason is the value of the linear near resonance frequencies, $f_{m}$'s, for the three pairs which are approximately, $\left[463,555\right]$, $\left[277,555\right]$, and $\left[25,126\right]$ respectively, where the numeric values inside the square brackets are in $\text{kHz}$. Such values of the resonance frequencies allow an examination of the linear Bjerknes theory if driving frequencies are in the range of $20-30\text{ kHz}$. The low driving frequencies also help ensure negligible time-delay between the bubbles which would otherwise further complicate the bubble-bubble interaction. The maximum time-delay for the first two pairs are approximately, $0.2\text{ }\%$ and $0.3\text{ }\%$ of the time-period of the sound field at $30 \text{ kHz}$. For the third pair, the maximum delay is close to $1.3\text{ }\%$ of the time-period of the sound field at $27 \text{ kHz}$. Also, the difference in size of the bubbles allow an investigation of the nonlinear coupling in bubble-pairs with different degree of asymmetricity as the ratio of equilibrium radii, $R_{10}/R_{20}$, for the three pairs are approximately, $1.2$, $2$, and $5$ respectively. Moreover, the bubbles from the first two pairs have a higher probability of formation within the assumed physical conditions, see Fig.~1 in \cite{Mettin1997}. 

As shown in Table~\ref{bubble_table}, the distance between the bubbles in the first two pairs is kept four times their combined equilibrium radii at pressure magnitudes of $0.7\text{ atm}$ and $1\text{ atm}$. The distance is increased to ten times their combined equilibrium radii at $1.3\text{ atm}$. This ensures that the separation between the bubbles is always greater than the combined radii of the two bubbles at any instant. The same approach is adopted in Ref.~\cite{Mettin1997} as well to avoid losing the sphericity of bubbles. But, in strong driving fields $(> 1 \text{ atm})$, sphericity is most probably lost. The resulting shape oscillations may also exchange energy with the volume oscillations, but the implications from such a coupling for micrometer size bubbles are negligible \cite{Huang2018,Zhang2018}. Besides, the bubble cavitation becomes transient if the separation distance is reduced to less than eight times their combined equilibrium radii under very strong driving fields, $P_s \ge 1.3\text{ atm}$. This is expected as bubbles may then grow into each other during their maximum growth phase which will immediately affect their stability and sphericity. This will also risk the generation of micro-jets if bubbles touched each other at opposite phases \cite{Han2015,Huang2018}. In such a transient cavitation collapse, a bubble often disintegrates into a collection of smaller bubbles. Further, depending on the driving conditions and the magnitude of the secondary Bjerknes force, the bubbles may coalesce together if the force is attractive \cite{Oguz1990,Jiao2015a}. It is evident that for problems involving bubbles separated by very small distances in very strong fields, the framework of KMEs may not be reliable. Rather, boundary element methods should be preferred \cite{Huang2018}. Moreover, since the focus of this paper is to study Bjerknes force in stable cavitation, effects like jetting, formation of capillary waves and shock waves that mostly occur during the violent phases of transient cavitation, are not pursued here. All bubble-pairs in this study exhibited stable cavitation, at least for five to ten wave cycles.  

A distinction between the near field and the far field that is also found relevant in this study is provided by Prosperetti in Ref.~\cite{Prosperetti1993}. The near field extends to a distance that is of the same order as the bubble radius. And, the far field scales as, $c\cdot \tau$. The far field results from the strong nonlinear coupling and the weak nonlinear coupling are almost identical. Therefore, they are neither shown, nor discussed here. Also, chaotic behavior was not observed in this study.

The numerical results for the secondary Bjerknes force are summarized in Table~\ref{bubble_table}. For the first two bubble pairs, $\left(6,5\right)$, and $\left(10,5\right)$, the weak nonlinear coupling always results into a net attractive force between the bubbles. This is in agreement with the linear Bjerknes theory as both bubbles are weakly driven below their resonance frequencies. But, strong coupling predicts both attractive and repulsive force. Also, strong nonlinear coupling gives a greater magnitude of the secondary Bjerknes force than that from the weak nonlinear coupling. The weaker the strength of the driving field, the larger the difference in magnitude of the forces is observed, often by more than two orders of magnitude. Further, the symmetry restrictions on weak coupling yield almost the same magnitude and the same sign of the mutual Bjerknes force, though minor differences can be observed in strong driving field of $1.3 \text{ atm}$. On the contrary, the strong coupling leads to both asymmetricity in magnitude as well as asymmetricity in sign reversal of the Bjerknes force. In the case of symmetric sign reversal, i.e., when both $\left\langle F_{12}\right\rangle$ and $\left\langle F_{21}\right\rangle$ are positive, both bubbles repel each other. In contrast, in the case of asymmetric sign reversal, one of the bubbles exerts an attractive force on the second bubble, but the second bubble exerts a repulsive force on the first bubble. An example of this is the first pair $\left(6,5\right)$ that is separated by a distance of $44\text{ }\mu m $, and is under excitation by a sound field of strength of $0.7\text{ atm}$, and frequency, $20\text{ kHz}$. The resulting values of $\left\langle F_{12}^{s}\right\rangle $ and $\left\langle F_{21}^{s}\right\rangle $ are $+1.4\text{ }\mu N $ and $-0.03\text{ }\mu N $ respectively. Such a situation would correspond to a \textit{bubble-bubble chase} like phenomena where the first bubble is attracted by the second bubble, but the latter is repelled by the former. In such a situation, the final behavior of the bubble-pair would be determined by the resultant, $\left\langle F_{12}^{s}\right\rangle + \left\langle F_{21}^{s}\right\rangle$  as shown in the last column of the Table~\ref{bubble_table}, which in this case will be a repulsion. Such a case of bubble-pair translation has recently been predicted using the boundary element method \cite{Huang2018}, but primary Bjerknes force was attributed for it. This is because the secondary Bjerknes force has traditionally been considered to yield either attraction, or, repulsion. Besides the asymmetricity in the sign reversal, large inequality in the magnitude of the mutual Bjerknes force is also observed in strong coupling. It should be emphasized that contrary to the possibility of asymmetricity in the Bjerknes force due to time delay \cite{Mettin1997} and viscous dissipation \cite{Pelekasis2004}, here it arises from the strong nonlinear coupling between the bubbles. Thus, a complete understanding of the strong coupling is required. Otherwise, the effects from the secondary Bjerknes force could be mistaken as that from the primary Bjerknes force, and vice-versa.

The bubbles in the third pair $\left(112,22\right)$, with a separation of $450\text{ }\mu m $, have been experimentally observed to attract each other when excited by a sound field of strength, $0.03\text{ atm}$, and frequency, $27\text{ kHz}$ \cite{Yoshida2011}. However, since the exciting frequency lies between the linear resonance frequencies of the two bubbles, the bubbles were expected to repel each other in accordance to the linear Bjerknes theory. Both weak coupling theory and strong coupling theory predict attractive Bjerknes force as documented in Table~\ref{bubble_table}. But, the observed repulsive force was much greater than the predictions from the linear theory, though the exact value was not mentioned. Since strong coupling yields a relatively larger magnitude of the Bjerknes force than that from the weak coupling, the findings may be seen as inclined in its favor. Further, it is reasonable to expect that bubbles with such a large difference in size may not always exert the same magnitude of the force on each other that is imposed by the symmetricity conditions of the weak nonlinear coupling. The outcome from the strong nonlinear coupling could have been validated with more authority, if the bigger bubble was not fixed in the experiment as then the strong attractive force exerted by the smaller bubble on the bigger bubble would have been observed.

\begin{table}
\centering{}%
\begin{tabular}{|c|c|c|c|c|c|c|}
\hline 
\begin{tabular}{c}
$\left(R_{10},\text{}R_{20}\right)$\tabularnewline
$\left(\mu m\right)$\tabularnewline
\end{tabular} & %
\begin{tabular}{c}
$d$\tabularnewline
$\left(\mu m\right)$\tabularnewline
\end{tabular} & %
\begin{tabular}{c}
$P_{s}$\tabularnewline
$\left(\text{atm}\right)$\tabularnewline
\end{tabular} & %
\begin{tabular}{c}
$f_{s}$\tabularnewline
$\left(\text{kHz}\right)$\tabularnewline
\end{tabular} & %
\begin{tabular}{c}
$\left\langle F_{12}\right\rangle $\tabularnewline
$\left(\mu N\right)$\tabularnewline
\end{tabular} & %
\begin{tabular}{c}
$\left\langle F_{21}\right\rangle $\tabularnewline
$\left(\mu N\right)$\tabularnewline
\end{tabular} & %
\begin{tabular}{c}
$\left\langle F_{12}\right\rangle +\left\langle F_{21}\right\rangle $\tabularnewline
$\left(\mu N\right)$\tabularnewline
\end{tabular}\tabularnewline
\hline 
\hline 
\multirow{9}{*}{$\left(6,5\right)$} & \multirow{6}{*}{$44$} & \multirow{3}{*}{$0.7$} & $20$ & $+1.4\text{ }\left(-0.00008\right)$ & $-0.03\text{ }\left(-0.00008\right)$ & $+1.37\text{ }\left(-0.00016\right)$\tabularnewline
\cline{4-7} \cline{5-7} \cline{6-7} \cline{7-7} 
 &  &  & $25$ & $+1.3\text{ }\left(-0.00013\right)$ & $-0.2\text{ }\left(-0.00013\right)$ & $+1.1\text{ }\left(-0.00026\right)$\tabularnewline
\cline{4-7} \cline{5-7} \cline{6-7} \cline{7-7} 
 &  &  & $30$ & $+1.25\text{ }\left(-0.00019\right)$ & $-0.25\text{ }\left(-0.00019\right)$ & $+1\text{ }\left(-0.00038\right)$\tabularnewline
\cline{3-7} \cline{4-7} \cline{5-7} \cline{6-7} \cline{7-7} 
 &  & \multirow{3}{*}{$1$} & $20$ & $+1.36\text{ }\left(-0.007\right)$ & $+0.01\text{ }\left(-0.007\right)$ & $+1.37\text{ }\left(-0.014\right)$\tabularnewline
\cline{4-7} \cline{5-7} \cline{6-7} \cline{7-7} 
 &  &  & $25$ & $+1.1\text{ }\left(-0.01\right)$ & $-0.5\text{ }\left(-0.01\right)$ & $+0.6\text{ }\left(-0.02\right)$\tabularnewline
\cline{4-7} \cline{5-7} \cline{6-7} \cline{7-7} 
 &  &  & $30$ & $+1.1\text{ }\left(-0.01\right)$ & $-0.7\text{ }\left(-0.01\right)$ & $+0.4\text{ }\left(-0.02\right)$\tabularnewline
\cline{2-7} \cline{3-7} \cline{4-7} \cline{5-7} \cline{6-7} \cline{7-7} 
 & \multirow{3}{*}{$110$} & \multirow{3}{*}{$1.3$} & $20$ & $-2.2\text{ }\left(-6.4\right)$ & $-6.4\text{ }\left(-5.5\right)$ & $-8.6\text{ }\left(-11.9\right)$\tabularnewline
\cline{4-7} \cline{5-7} \cline{6-7} \cline{7-7} 
 &  &  & $25$ & $+5.7\text{ }\left(-3.2\right)$ & $-2.3\text{ }\left(-3.7\right)$ & $+3.4\text{ }\left(-6.9\right)$\tabularnewline
\cline{4-7} \cline{5-7} \cline{6-7} \cline{7-7} 
 &  &  & $30$ & $-4\text{ }\left(-1.9\right)$ & $-7.2\text{ }\left(-1\right)$ & $-11.2\text{ }\left(-2.9\right)$\tabularnewline
\hline 
\hline 
\multirow{9}{*}{$\left(10,5\right)$} & \multirow{6}{*}{$60$} & \multirow{3}{*}{$0.7$} & $20$ & $+0.03\text{ }\left(-0.00024\right)$ & $+0.14\text{ }\left(-0.00024\right)$ & $+0.17\text{ }\left(-0.00048\right)$\tabularnewline
\cline{4-7} \cline{5-7} \cline{6-7} \cline{7-7} 
 &  &  & $25$ & $-0.06\text{ }\left(-0.0004\right)$ & $-0.26\text{ }\left(-0.0004\right)$ & $-0.32\text{ }\left(-0.0008\right)$\tabularnewline
\cline{4-7} \cline{5-7} \cline{6-7} \cline{7-7} 
 &  &  & $30$ & $-0.1\text{ }\left(-0.00052\right)$ & $-0.4\text{ }\left(-0.00061\right)$ & $-0.5\text{ }\left(-0.00113\right)$\tabularnewline
\cline{3-7} \cline{4-7} \cline{5-7} \cline{6-7} \cline{7-7} 
 &  & \multirow{3}{*}{$1$} & $20$ & $-0.4\text{ }\left(-0.03\right)$ & $+0.05\text{ }\left(-0.03\right)$ & $-0.35\text{ }\left(-0.06\right)$\tabularnewline
\cline{4-7} \cline{5-7} \cline{6-7} \cline{7-7} 
 &  &  & $25$ & $+0.6\text{ }\left(-0.03\right)$ & $-1\text{ }\left(-0.03\right)$ & $-0.4\text{ }\left(-0.06\right)$\tabularnewline
\cline{4-7} \cline{5-7} \cline{6-7} \cline{7-7} 
 &  &  & $30$ & $+0.9\text{ }\left(-0.03\right)$ & $-1.7\text{ }\left(-0.03\right)$ & $-0.8\text{ }\left(-0.06\right)$\tabularnewline
\cline{2-7} \cline{3-7} \cline{4-7} \cline{5-7} \cline{6-7} \cline{7-7} 
 & \multirow{3}{*}{$150$} & \multirow{3}{*}{$1.3$} & $20$ & $-3.3\text{ }\left(-4\right)$ & $-10.2\text{ }\left(-6.6\right)$ & $-13.5\text{ }\left(-10.6\right)$\tabularnewline
\cline{4-7} \cline{5-7} \cline{6-7} \cline{7-7} 
 &  &  & $25$ & $+7\text{ }\left(-2.1\right)$ & $-4.3\text{ }\left(-3.4\right)$ & $+2.7\text{ }\left(-5.5\right)$\tabularnewline
\cline{4-7} \cline{5-7} \cline{6-7} \cline{7-7} 
 &  &  & $30$ & $-3.9\text{ }\left(-1.1\right)$ & $-8.4\text{ }\left(-2.1\right)$ & $-12.3\text{ }\left(-3.2\right)$\tabularnewline
\hline 
\hline 
\multirow{9}{*}{$\left(112,22\right)$} & \multirow{6}{*}{$450$} & \multirow{3}{*}{$0.03$} & $20$ & $-2.3\text{ }\left(-0.003\right)$ & $-21.7\text{ }\left(-0.003\right)$ & $-24\text{ }\left(-0.006\right)$\tabularnewline
\cline{4-7} \cline{5-7} \cline{6-7} \cline{7-7} 
 &  &  & $25$ & $-2.9\text{ }\left(-0.02\right)$ & $-26.2\text{ }\left(-0.01\right)$ & $-29.1\text{ }\left(-0.03\right)$\tabularnewline
\cline{4-7} \cline{5-7} \cline{6-7} \cline{7-7} 
 &  &  & $27$ & $-3\text{ }\left(-0.02\right)$ & $-27\text{ }\left(-0.01\right)$ & $-30\text{ }\left(-0.03\right)$\tabularnewline
\cline{3-7} \cline{4-7} \cline{5-7} \cline{6-7} \cline{7-7} 
 &  & \multirow{3}{*}{$0.1$} & $20$ & $-2.1\text{ }\left(-0.02\right)$ & $-21.5\text{ }\left(-0.02\right)$ & $-23.6\text{ }\left(-0.04\right)$\tabularnewline
\cline{4-7} \cline{5-7} \cline{6-7} \cline{7-7} 
 &  &  & $25$ & $-2.8\text{ }\left(-0.1\right)$ & $-26.3\text{ }\left(-0.06\right)$ & $-29.1\text{ }\left(-0.16\right)$\tabularnewline
\cline{4-7} \cline{5-7} \cline{6-7} \cline{7-7} 
 &  &  & $27$ & $-3\text{ }\left(-0.07\right)$ & $-27\text{ }\left(-0.06\right)$ & $-30\text{ }\left(-0.13\right)$\tabularnewline
\cline{2-7} \cline{3-7} \cline{4-7} \cline{5-7} \cline{6-7} \cline{7-7} 
 & \multirow{3}{*}{$700$} & \multirow{3}{*}{$0.4$} & $20$ & $-0.7\text{ }\left(-0.1\right)$ & $-9.2\text{ }\left(-0.1\right)$ & $-9.9\text{ }\left(-0.2\right)$\tabularnewline
\cline{4-7} \cline{5-7} \cline{6-7} \cline{7-7} 
 &  &  & $25$ & $-1.3\text{ }\left(-0.17\right)$ & $-9.2\text{ }\left(-0.17\right)$ & $-10.5\text{ }\left(-0.34\right)$\tabularnewline
\cline{4-7} \cline{5-7} \cline{6-7} \cline{7-7} 
 &  &  & $27$ & $-1.4\text{ }\left(-0.1\right)$ & $-8.1\text{ }\left(-0.15\right)$ & $-9.5\text{ }\left(-0.25\right)$\tabularnewline
\hline
\end{tabular}\caption{Comparison of secondary Bjerknes force obtained from strong nonlinear coupling and weak nonlinear coupling. The values inside the parentheses in the last three columns correspond to those obtained from the weak nonlinear coupling.}
\label{bubble_table}
\end{table}

\begin{figure}
\begin{centering}
\includegraphics[width=1.04\columnwidth]{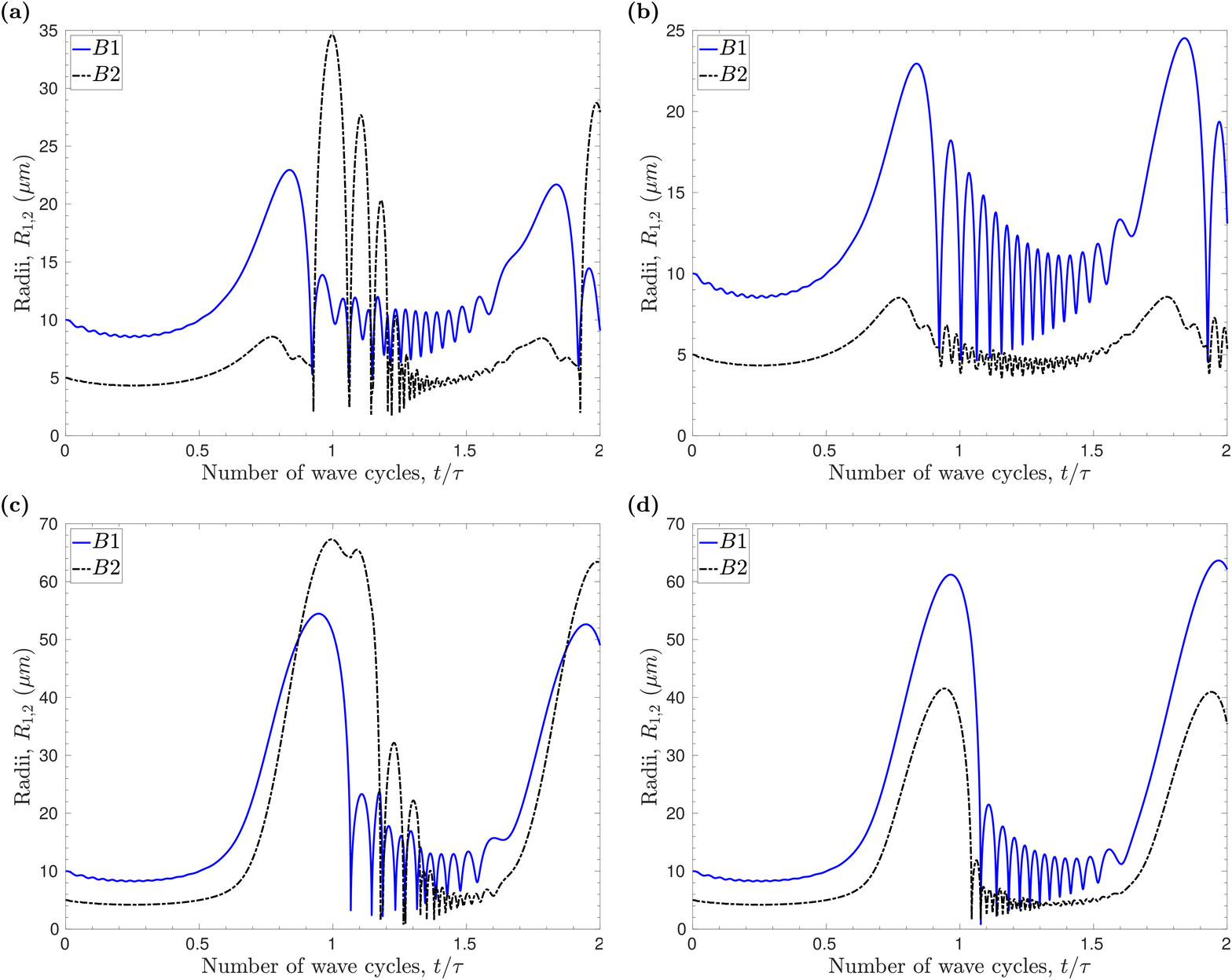}
\par\end{centering}
\caption{(Color online) Radii evolution for the bubble-pair; $B1$ (solid line)
and $B_{2}$ (dash-dotted line), with $R_{10}=10\text{ }\mu m$ and
$R_{20}=5\text{ }\mu m$, for the first two wave cycles of frequency, $f_{s}=20\text{ kHz}$. Subfigures (a) and (c) correspond to strong nonlinear coupling.
And, subfigures (b) and (d) correspond to weak nonlinear coupling.
The top and bottom rows represent radii evolution at $P_{s}=1\text{ atm}$,
$d=75\text{ }\mu m$, and $P_{s}=1.3\text{ atm}$, $d=150\text{ }\mu m$,
respectively.}
\centering
\label{All_Radii}
\end{figure}

The radial pulsations of the second bubble pair, $\left(10,5\right)$, is shown in Fig.~\ref{All_Radii}. The critical Blake radii \cite{Harkin1999} for the bubbles are approximately $48 \text{ } \mu m$ and $18\text{ } \mu m$ respectively. It can be seen from Fig.~\ref{All_Radii}(b) that the two bubbles under weak nonlinear coupling at $1 \text{ atm}$, did not grew larger than their respective critical Blake radius, implying stable cavitation. But, as shown in the Fig.~\ref{All_Radii}(a), the smaller bubble, $B2$, surpassed its critical Blake threshold at the same driving pressure due to the strong coupling affects from $B1$. In contrast, as shown in Figs.~\ref{All_Radii}(c) and (d), both bubbles grew past their respective critical radii because the driving field is sufficiently strong at $1.3 \text{ atm}$. Bubbles that experience a growth larger than their critical Blake threshold are known to exhibit nonlinear resonance \cite{Mettin1997}, that is observed in Figs.~\ref{All_Radii}(a), (c), and (d). The physical mechanism underlying such an explosive growth is an interplay between the liquid negative pressure and the controlling effects of the surface tension. Further, since surface tension dictates the growth of small cavitation bubbles even more in weak acoustic fields, it may be difficult to observe nonlinear resonance growth in such cases.

The dramatic growth of the smaller bubble, $B2$, beyond its critical Blake radius can be explained using the concept of dynamical Blake threshold \cite{Mettin1997,Louisnard2011}. As seen in Fig.~\ref{All_Radii}(a), as the larger bubble, $B1$, undergoes its first major collapse just before the end of the first wave cycle, it pulls the fluid surrounding its boundary inward. This convective contribution from the $B1$ that is best taken care in the framework of strong coupling lowers the liquid pressure surrounding the smaller bubble, $B2$. This facilitates an explosive growth of $B2$ which due to its smaller size easily gets excited even more. This effect is more pronounced either when the bubbles are closely-spaced, or when they are placed in a stronger driving field as shown in Fig.~\ref{All_Radii}(c). It can be seen from Figs.~\ref{All_Radii}(a) and \ref{All_Radii}(c) that the main nonlinear resonance jump of $B2$ occurs approximately at the same time when $B1$ undergoes contraction. This aspect is almost absent in Figs.~\ref{All_Radii}(b) and \ref{All_Radii}(d) that results from the weak coupling. Surprisingly, this is in agreement with the findings from Ref.~\cite{Ida2009b} which describes a reduction of the effective cavitation threshold pressure for a smaller bubble that is in the proximity of a larger bubble. Further as seen in Fig.~\ref{All_Radii}(a), a few successive oscillations of $B1$ after its first major collapse have relatively lesser magnitude when compared to its respective oscillations from Fig.~\ref{All_Radii}(b). This is because of the explosive growth of $B2$ that pushes the fluid outward, which in turn, increases the liquid pressure surrounding $B1$, and hence restrains its growth. But, since the convective flow of the fluid is not included in the case of weak coupling, the critical pressure thresholds for the bubbles are not altered.

It should be noted that bubble dynamics after such nonlinear resonance jumps switches drastically from quasi-static oscillations to rapid phases of growth and violent collapses. This is because after the bubble reaches its maximum growth, the internal pressure in the bubble drops so low that it is unable to retain the inward flow of the surrounding liquid. Thereafter, the bubble may also enter a transient phase in which it collapses violently and undergoes rapid successions of afterbounces. The collapse velocities of the bubbles, $\dot{R}_{1,2}$, have been less than $1 \text{ }m/s$, $10 \text{ }m/s$, and $100 \text{ }m/s$ at $0.7 \text{ atm}$, $1 \text{ atm}$, and $1.3 \text{ atm}$ respectively. Also, the bubbles have a relatively greater collapse velocity in the case of strong coupling. Therefore, compressibility effects should be taken into account for closely-spaced bubble-pairs in stronger driving fields. Furthermore, since the ratio, $f_s/f_m$, is not an integer, it may lead to ultraharmonic resonances that may get even more pronounced in the case of strong nonlinear coupling \cite{Louisnard2011}.

\begin{figure}
\begin{centering}
\includegraphics[width=1.04\columnwidth]{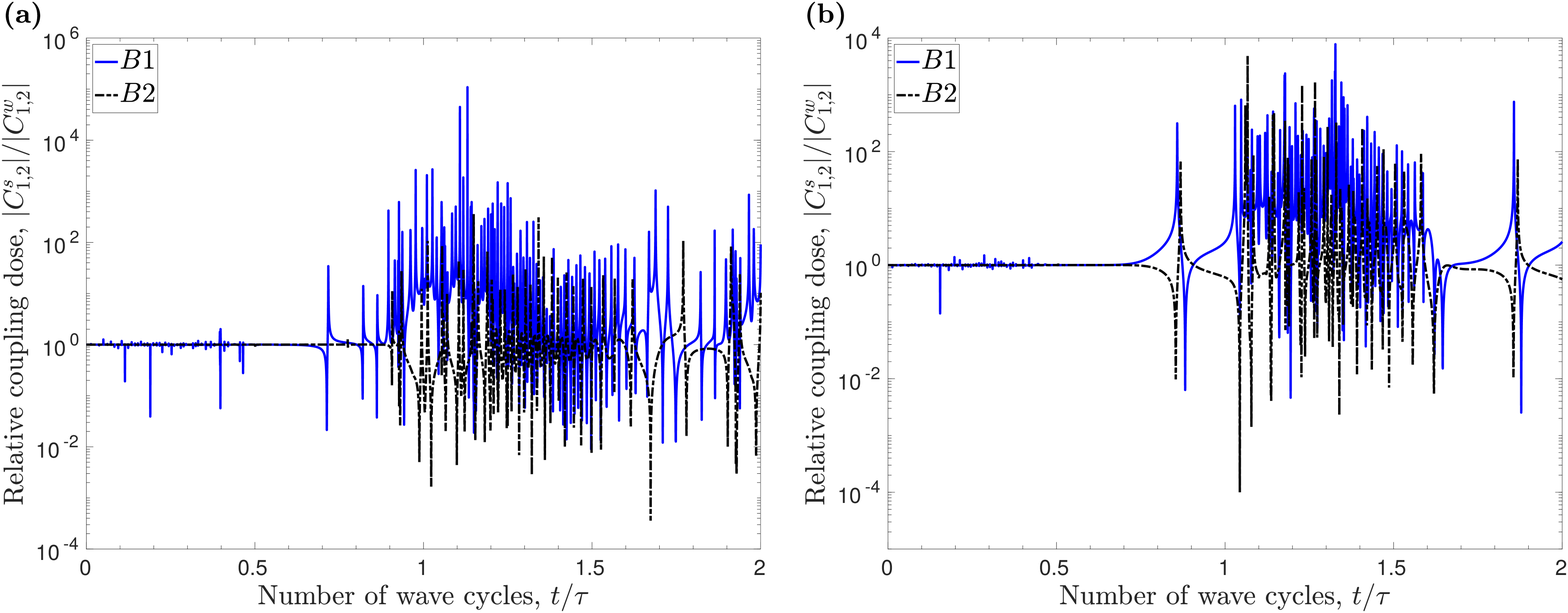}
\par\end{centering}
\caption{(Color online) Ratio of coupling dose from strong nonlinear coupling
and weak nonlinear coupling for the bubble-pair; $B1$ (solid line), and
$B_{2}$ (dash-dotted line), with $R_{10}=10\text{ }\mu m$ and $R_{20}=5\text{ }\mu m$, for the first two wave cycles of frequency, $f_{s}=20\text{ kHz}$. Subfigures (a)
and (b) correspond to the conditions, $P_{s}=1\text{ atm}$, $d=75\text{ }\mu m$,
and $P_{s}=1.3\text{}\text{ atm}$, $d=150\text{ }\mu m$, respectively.}
\begin{centering}
\label{All_Coupling}
\par\end{centering}
\end{figure}

Another important observation from Fig.~\ref{All_Radii} is that effects from strong coupling and weak coupling are almost unnoticeable until seventy percent of the first wave cycle. This can also be confirmed from Figs.~\ref{All_Coupling}(a) and (b), where the ratio of the absolute values of strong coupling dose and weak coupling dose are plotted for the two bubbles. Surprisingly, since the ratio between the doses for the smaller bubble,
$B2$, is less than one at multiple instances after the first wave cycle, this means that the weak nonlinear coupling is not necessarily lesser in magnitude. But, rather it over-predicts the coupling dose for $B2$. The opposite happens for the bigger bubble, $B1$, for which the weak nonlinear theory under-predicts the coupling dose. 

\begin{figure}
\begin{centering}
\includegraphics[width=1.04\columnwidth]{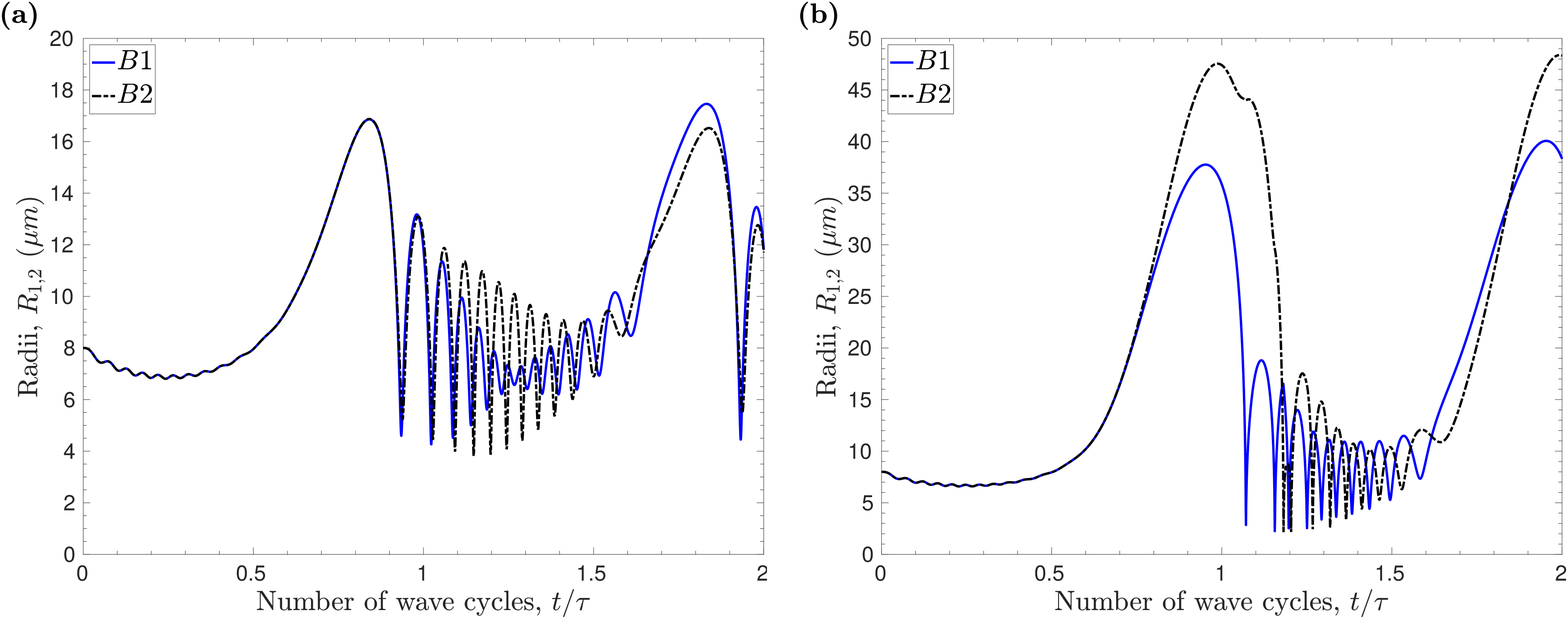}
\par\end{centering}
\caption{(Color online) Radii evolution in strong nonlinear coupling for the bubble-pair; $B1$ (solid line)
and $B_{2}$ (dash-dotted line), with $R_{10}=R_{20} = 8\text{ }\mu m$, separated by a distance of $128\text{ } \mu m$,  for the first two wave cycles of frequency, $f_{s}=30\text{ kHz}$. Subfigures (a) and (b) correspond to the conditions, $P_{s}=1\text{ atm}$, and $1.3\text{ atm}$ respectively.}
\begin{centering}
\label{All_Comparison}
\par\end{centering}
\end{figure}

The effects of the strong nonlinearity in coupling is best witnessed in the case of the symmetric bubble-pair, $\left(8,8\right)$. As shown in Fig.~\ref{All_Comparison}, the radii-evolution of the bubbles that are separated by a distance eight times their combined equilibrium radii, do not look identical even if their equilibrium radii is the same, i.e., ${R}_{10}={R}_{20} = 8 \text{ }\mu m$. As explained in the paragraph following Eq.~$\left(\ref{asymmetric_Bjerknes}\right)$, the strong nonlinear coupling hinders the equality of the radial velocities even when the bubbles are of the same size. As shown in Fig.~\ref{All_Comparison}(b), both bubbles exhibit nonlinear resonance jump beyond their critical Blake radius of $35 \text{ } \mu m$ at $1.3 \text{ atm}$. Interestingly, the jump is not the same for both bubbles. The respective plots from the weak nonlinear coupling are not shown since the curves for the two bubbles then essentially coincide. Further, in light of the linear theory approach from Ref.~\cite{Lanoy2015}, the resonance frequencies for the symmetric and asymmetric modes for this bubble-system are suppose to be identical. Clearly, the observations from Fig.~\ref{All_Comparison} are against the predictions from the linear theory as well as from the weak nonlinear theory. Interestingly, the divergence of the resonance frequencies due to radiation coupling in closely spaced identical bubbles has already been explained by Feuillade \cite{Feuillade1995}.

The bubble pair, $\left(8,8\right)$, shown in Fig.~\ref{All_Comparison} yields secondary Bjerknes force, $\left\langle F_{12}^{s}\right\rangle = +0.17 \text{ }\mu N$ and  $\left\langle F_{12}^{s}\right\rangle = -0.18  \text{ }\mu N$, at $P_{s}=1\text{ atm}$, and $f_s = 30 \text{ kHz}$. The respective values of the forces at $1.3\text{ atm}$ are, $-1.33 \text{ } \mu N$, and $-5.1 \text{ } \mu N$, respectively. In the case of weak nonlinear coupling, $\left\langle F_{12}^{w}\right\rangle = \left\langle F_{21}^{w}\right\rangle =-0.07\text{ } \mu N$ at $P_{s}=1\text{ atm}$, and $\left\langle F_{12}^{w}\right\rangle = \left\langle F_{21}^{w}\right\rangle =-2.6\text{ } \mu N$ at $P_{s} = 1.3\text{ atm}$. In the case of strong coupling, the same pair when separated by just two times their combined equilibrium radii exerts Bjerknes force of approximately, $\left\langle F_{12}^{s}\right\rangle=+3.1 \text{ } \mu N$, and $\left\langle F_{21}^{s}\right\rangle=-2 \text{ }\mu N$, for all three magnitudes of the driving pressure, $0.1\text{ atm}$, $0.05 \text{ atm}$, and $0.01 \text{ atm}$. Thus at such weak driving fields, strong nonlinear coupling predicts a net repulsive force between the bubbles. On the contrary, weak coupling predicts an attractive force that is of the order of nano-Newton or less. This is interesting since formation of bubble-grapes has been observed when bubbles separated by distances comparable to their equilibrium sizes are excited by weak acoustic fields as low as $0.035 \text{ atm}$ \cite{Doinikov1995a,Doinikov1995b,Ida2003}.
 
\section{Conclusion\label{sec:Conclusion}}

The implications of strong nonlinear coupling in cavitation bubble-pairs have been investigated. It is found that the bubble-bubble interaction involves two mechanisms. First, the regular radiation coupling between the bubbles. Second, the interaction mediated by the fluid flow that arises predominantly in the near field due to the volume pulsations of the bubbles. Strong nonlinear coupling affects closely spaced bubble pairs; both symmetric and asymmetric. The convective contribution from the strong nonlinear coupling adds to the attraction between the bubbles, but opposes their repulsion. The results obtained from the strong coupling are different than those from the weak coupling in weak, regular, and strong acoustic fields. It is found that if strong nonlinear coupling is taken into account, it leads to an asymmetricity in the magnitude as well as in the sign of the secondary Bjerknes force. It is also envisioned that the strong nonlinear coupling led sign reversal of the secondary Bjerknes force in weak acoustic fields may play a role in the formation of stable bubble clusters. Interestingly, the inherent asymmetric nature of the Bjerknes force may possibly explain why theories motivated from the study of pulsating spheres did not succeed in describing the symmetric laws of gravitation and electromagnetism \cite{Crum1975}. 

Further, the nonlinear resonance jumps observed for the smaller bubble in strong acoustic fields is attributed to the lowering of its dynamical Blake threshold pressure. This occurs due to the convective fluid flow from the volume pulsations of the bigger bubble. Such nonlinear resonance jumps are observed even in the case of symmetric bubble-pairs, implying bubbles of the same size may not share the same resonance frequency if they are strongly coupled. It is expected that the results reported here may be useful in industrial applications where a manipulation of bubble clusters to control cavitation effects is required. A quantitative analysis of the divergence of the resonance frequencies, the lowering of the critical threshold pressure, and the presence of charges, in strongly coupled bubble-pairs are intended for future work. 

\section*{Acknowledgment}

The derivation part of this manuscript was partly completed during the author's postdoctoral stay in ARCEx at UiT The Arctic University of Norway in Troms\o. The author would therefore like to acknowledge the funding provided by ARCEx (Research Council of Norway Grant No. 228107) and Statoil\textquoteright s ``Akademia'' agreement.

\bibliographystyle{apsrev4-1.bst}
\bibliography{StrongNonlinearBubbleCoupling.bib}

\vspace{20pt}
\textcolor{blue}{\large \textbf{Please cite to this manuscript as published in:}}\vspace{0pt}
\\

\large V.~Pandey, “Asymmetricity and sign reversal of secondary Bjerknes force from strong nonlinear coupling in cavitation bubble pairs,” Phys.~Rev.~E~\textbf{99}, 042209 (2019).\vspace{0pt}
\\

 \href{https://doi.org/10.1103/PhysRevE.99.042209}{\textbf{DOI: https://doi.org/10.1103/PhysRevE.99.042209}}

\end{document}